\documentclass[sigconf,screen]{acmart}

\AtBeginDocument{%
  }

\usepackage{xcolor}
\usepackage{graphicx}
\usepackage{tabularx}
\usepackage{graphicx}
\usepackage{booktabs} 
\usepackage{xspace}
\usepackage{fancyvrb}
\usepackage{balance}
\usepackage[ruled,linesnumbered,boxed]{algorithm2e} 
\usepackage{listings}
\usepackage{url}
\usepackage{multirow, array}
\usepackage{colortbl}
\usepackage{pifont}
\usepackage{booktabs}
\usepackage{tikz}
\usepackage{stfloats}
\usepackage{float}
\usepackage{adjustbox}
\usepackage{hhline}
\usepackage{mdframed}

\makeatletter
\let\@authorsaddresses\@empty
\makeatother

\makeatletter
\def\@ACM@checkaffil{
    \if@ACM@instpresent\else
    \ClassWarningNoLine{\@classname}{No institution present for an affiliation}%
    \fi
    \if@ACM@citypresent\else
    \ClassWarningNoLine{\@classname}{No city present for an affiliation}%
    \fi
    \if@ACM@countrypresent\else
        \ClassWarningNoLine{\@classname}{No country present for an affiliation}%
    \fi
}
\makeatother

\newcommand{\tool}{\textsc{Wingfuzz}\xspace}

\newcommand{\pg}{PostgreSQL\xspace}

\newcommand{\monetdb}{MonetDB\xspace}
\newcommand{\duckdb}{DuckDB\xspace}
\newcommand{\clickhouse}{ClickHouse\xspace}

\newcommand{\mytilde}{\raise.17ex\hbox{$\scriptstyle\mathtt{\sim}$}}

\definecolor{trolleygrey}{rgb}{0.38, 0.38, 0.38}

\definecolor{darkgreen}{RGB}{30,180,30}

\newcommand{\llm}{LLM}

\newcounter{reccounter}[subsection] 
\renewcommand{\thereccounter}{\texttt{REC} \texttt{\arabic{subsection}.\arabic{reccounter}}~\xspace} 

\newcommand{\recstep}{%
    \refstepcounter{reccounter}%
    \thereccounter%
}
\begin{document}

\title{When Fuzzing Meets LLMs: Challenges and Opportunities}


\author{Yu Jiang*, Jie Liang*, Fuchen Ma*, Yuanliang Chen*, Chijin Zhou*, Yuheng Shen*}
\author{Zhiyong Wu*, Jingzhou Fu*, Mingzhe Wang*, ShanShan Li\textsuperscript{\dag}, Quan Zhang*}
\affiliation{%
  \institution{*School of Software, Tsinghua University. \textsuperscript{\dag} National University of Defense Technology.}
}

\setlength{\intextsep}{0.1\baselineskip plus 0.1\baselineskip minus 0.1\baselineskip}
\setlength{\abovecaptionskip}{0.2\baselineskip plus 0.1\baselineskip minus 0.1\baselineskip}
\setlength{\belowcaptionskip}{0.01\baselineskip plus 0.1\baselineskip minus 0.1\baselineskip}
\setlength{\textfloatsep}{0.1\baselineskip plus 0.1\baselineskip minus 0.1\baselineskip}

\renewcommand{\shortauthors}{Jiang et al.}



\begin{abstract}
Fuzzing, a widely-used technique for bug detection, has seen advancements through Large Language Models (LLMs). 
Despite their potential, LLMs face specific challenges in fuzzing.
In this paper, we identified five major challenges of LLM-assisted fuzzing.
To support our findings, we revisited the most recent papers from top-tier conferences, conﬁrming that these challenges are widespread. 
As a remedy, we propose some actionable recommendations to help improve applying LLM in Fuzzing and conduct preliminary evaluations on DBMS fuzzing.
The results demonstrate that our recommendations effectively address the identified challenges.
\end{abstract}

\maketitle

\section{Introduction}
Fuzzing is a promising technique for software bug detection~\cite{liang2018fuzzing, chen2018systematic}.
Large Language Models (LLM) are rapidly gaining popularity across various applications for their versatility and capability~\cite{hadi2023large,hou2023large}. 
From natural language processing~\cite{liu2023evaluating, kumar2023large, chang2023survey} to code generation~\cite{li2023large, ji2023benchmarking}, LLM's broad utility is making it a prominent and sought-after solution in diverse domains.
This development has naturally influenced fuzzing research:
to help improve the fuzzing effectiveness, LLM has now become one of the key enablers to assist the core processes of fuzzing, including driver synthesis~\cite{zhang2023understanding, lyu2023prompt}, input generation~\cite{dantas2023large, dakhel2023effective}, and bug detection~\cite{ibrahimzada2023automated, deng2023large}.

While excelling in natural language analysis, LLM encounters some common pitfalls like limited context length~\cite{kaddour2023challenges} and hallucination problems~\cite{huang2023survey, lee2018hallucinations, rohrbach2018object}, etc. Consequently, LLM exhibits limitations in complex program analysis.
These pitfalls of LLM affect the effectiveness of fuzzing, leading to testing performance degradation, manifesting as high false positives, low test coverage, and limited scalability.


In this paper, we identify five common challenges when using LLM-based fuzzing technology: 
1) Firstly, they often produce low-quality outputs in fuzzing driver synthesis, lacking the precision required for effective bug detection.
2) Secondly, these models demonstrate a limited scope in their understanding and processing capabilities, constraining their utility in diverse fuzzing scenarios.
3) Thirdly, LLMs struggle with generating sufficiently diverse inputs during the fuzzing process, which is critical for thorough and effective bug detection.
4) Fourthly, they face challenges in maintaining the validity of generated inputs, a crucial factor for accurate and reliable fuzzing.
5) Lastly, LLMs' inaccurate understanding of bug detection mechanisms hinders their ability to identify and address complex software vulnerabilities effectively, thereby limiting their overall effectiveness in the fuzzing process.
We performed a comprehensive survey and revisited most recent fuzzing works that rely on LLM for tackling different problems in the fuzzing process. To our surprise, the results show that each work encounters at least one of these challenges.~\footnote{Remark: The purpose of this work is not to point fingers
or critique. Instead, it wants to show how we can overcome the challenges of LLM-assisted fuzzing and effectively leverage the advantages of LLMs and make it truly
beneficial for the fuzzing process.}

Although LLMs are widespread, it is more important for us to avoid its weakness, and at the same time take advantage of its strengths. To this end, we perform an impact analysis of the implications in three key fuzzing steps. These ﬁndings inspire us with some opportunities for better usage of LLM in each fuzzing step according to whether the corresponding corpus and documentation are rich.
Furthermore, we performed some preliminary evaluations according to these opportunities by applying LLM in fuzzing database management systems(DBMS).
The results demonstrate that the reasonable instantiation of those recommendations can overcome the challenges in LLM-assisted DBMS fuzzing.

\section{Challenges and Opportunities} 
\begin{figure}[!htp]
    \centering
    \includegraphics[width=\linewidth]{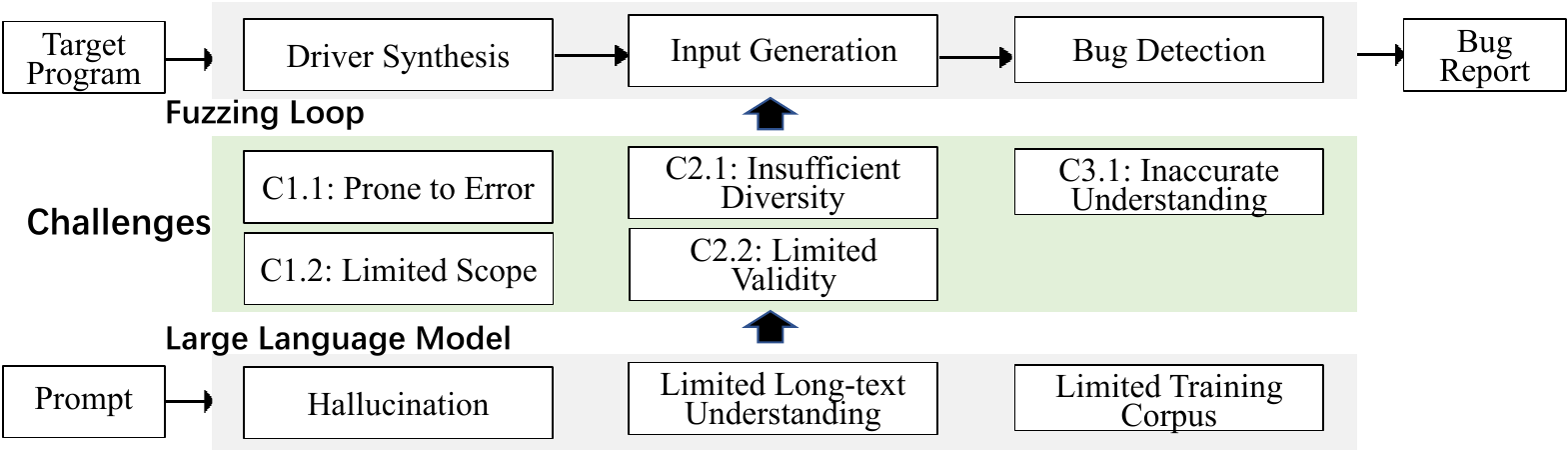}
    \vspace{0.1cm}
    \caption{Fuzzing Workflow with LLM enhanced.} 
    \label{fig:challenge}
\end{figure}

Despite that \llm{} have achieved great success, the application of \llm{} in fuzzing is often prone to several problems, ranging from deduction accuracy to adapt scalability.
Overlooking these issues may result in poor seed quality or omitting critical bugs, leading to a limited fuzzing performance.
In this section, we summarize the five challenges that commonly occur when applying \llm{} in fuzzing.
While these challenges might initially appear straightforward, they usually stem from small shortcomings that are typical in fuzzing.
We group these challenges with respect to the states of a typical fuzzing workflow, as depicted in ~\autoref{fig:challenge}.
\\


\subsection{Driver Synthesis}
\textbf{\indent Description.}
Recently, several pioneer works have been proposed to utilize LLMs to enhance driver synthesis~\cite{zhang2023understanding, lyu2023prompt, yang2023kernelgpt,deng2023large,deng2023large2}. Their basic idea is to use API documentation as the prompt context, and then ask LLMs to generate API invoking sequences as fuzzing drivers. For example, both TitanFuzz~\cite{deng2023large} and PromptFuzz~\cite{lyu2023prompt} design customized prompt templates to guide LLMs in generating code that follows programming syntax and semantics.

\textbf{Challenges.}
The application of LLMs to driver synthesis can be ineffective if done directly, as LLMs have a tendency to produce hallucinations~\cite{chang2023survey,kaddour2023challenges} and perform less effectively on programs that are not included in their training corpus~\cite{kaddour2023challenges}. 
These limitations present two challenges for driver synthesis.
The first one is that the synthesized drivers are \underline{\textit{prone to error}}, leading to a non-negligible number of false positives during fuzzing. 
For example, according to comprehensive evaluation on LLM-based driver synthesis for OSS-Fuzz projects~\cite{zhang2023understanding}, GPT-4 can correctly generate roughly 40\% drivers, while the rest of the drivers contain errors. 
Among the erroneous drivers, 93\% exhibit one or more of the following issues: type errors, mis-initialized function arguments, usage of non-existing identifiers, and imprecise control-flow dependencies. This occurrence primarily arises due to LLMs relying on pre-trained knowledge for driver synthesis, leading to the production of hallucinations~\cite{huang2023survey}.
The second challenge is that the application of directly using LLMs for driver synthesis has \underline{\textit{limited scope}} because LLMs have limited knowledge on unseen programs.
For those target programs, LLMs sometimes use training knowledge to fill the gap, thus generating incorrect API invoking sequences. 
For example, developers from Google's OSS-Fuzz project~\cite{google_ossfuzz_driver} attempted to leverage LLMs to synthesize drivers. Out of 31 tested OSS-Fuzz projects, 14 successfully compiled new targets and increased coverage with the synthesized drivers. The drivers unsuccessfully synthesized by LLMs typically originated from less common projects like \texttt{krb5} and \texttt{rtpproxy}. In contrast, LLMs are more likely to generate compilable and effective drivers for more common projects, such as \texttt{tinyxml2} and \texttt{cjson}.

\textbf{Recommendations.}  We have the following recommendations:

%
\recstep Some targets whose code or use cases have been included in the training corpus. \textit{For these cases, employing LLM for automated synthesis of fuzz drivers, complemented by error-guided corrective measures, is a practical approach.}  \label{rec:driver1}
Iteratively querying the \llm{} based on identified errors and fixing the errors are practical measures~\cite{zhang2023understanding}, which helps to address the \textit{prone-to-error} challenge.

For example, \texttt{libpng} is a common library and has already been seen by GPT4 in its training process. Consequently, it is possible to directly ask GPT4 to generate a fuzz testing driver for \texttt{libpng} by giving the prompt ``Generating LLVMFuzzerTestOneInput for test libpng.''
However, the generated driver might still contain errors in grammar or encounter issues during the process of compiling and linking. 
Test engineers can subsequently submit individual \llm{} queries containing the error messages to rectify these issues, occasionally necessitating multiple iterations.

\recstep For targets without a dedicated corpus in training, one can collect valuable materials such as function prototypes, example programs, or connection rules between functions. 
\textit{Conducting prompt engineering which involves embedding these materials, enhances the precision in generating logical sequences of function calls for the creation of drivers.} \label{rec:driver2}
The prompt engineering approach is a practical solution to tackle the challenge of \textit{limited scope}.

For example, \texttt{typst} is a new markup-based typesetting system like LaTex and claims it is more 
easier to learn and use.
To generate a fuzz driver for it, feed the prompt ``Generate LLVMFuzzerTestOneInput for typst'' to ChatGPT-3.5 will encounter hallucination problems and generate a completely non-existent driver.
Instead, the project \texttt{typst} has lots of documents and unit tests. 
Feeding these materials that illustrate the usage of the functions is helpful for \llm{}s to generate effective drivers~\cite{google_ossfuzz_driver}.
Additionally, it is also feasible to iteratively query \llm{}s to address any errors that may be present in the drivers.


\recstep Sometimes, even with adequate documentation and examples, \llm{}s can still encounter challenges in generating valid drivers at times, especially for extremely complex targets like Linux kernel. 
These systems frequently involve intricate dependencies among their APIs, or there exist implicit dependencies among lower-level systems that pose challenges for LLM to capture.
\textit{For these targets, it is advisable to refrain from relying on \llm{}s. Instead, it is more practical and feasible to explore conventional methods.} \label{rec:driver3}

For example, KSG~\cite{sun2022ksg} uses the ebpf to dynamically infer the kernel's system call argument type and value constraints. 
In contrast, LLM-based approaches such as KernelGPT~\cite{yang2023kernelgpt} use static inference based on kernel man pages and source code.
But they may find some complex dummy operations.
And it's hard for them to deduct pointer references.
Therefore, KSG can generate 2,433 Syzlang, which is $17.86\times$ more compared to KernelGPT~\cite{yang2023kernelgpt}.









\subsection{Input Generation}

\textbf{\indent Description.}
Recently, several pioneer works~\cite{ackerman2023large,xia2023universal,tamminga2023utilizing,yang2023white} have been proposed to utilize LLM to enhance input generation. Their basic idea is to use input specifications and input examples as the prompt context and then ask LLMs to generate new inputs.
For example, LLMFuzzer~\cite{ackerman2023large} feeds input specifications to LLMs to generate initial seeds for mutation-based fuzzers.

\textbf{Challenges.}
The application of LLMs to input generation can be ineffective if done directly, as LLMs heavily rely on training corpus and have limited long-text understanding~\cite{kaddour2023challenges,shaham2023zeroscrolls}. These limitations present two challenges for input generation.
The first one is that the generated inputs have \underline{\textit{insufficient diversity}}, leading to inefficient exploration of the input space.
This is because LLMs are pre-trained models and prone to responding to users' queries in a similar manner when given the same prompt context. Therefore, it is difficult for LLMs to generate diverse inputs if they only provide limited information. 
For example, ChatAFL~\cite{chatafl} demonstrates a significant limitation when directly applying LLMs to the RTPS protocol fuzzing. If only a limited amount of protocol information is provided in the prompts, LLMs can only generate inputs that cover 4 states out of 10 states that the RTPS protocol supported. This results in a substantial portion of the RTSP state remaining unexplored.
The second challenge is that the generated inputs often have \underline{\textit{limited validity}}, leading to early termination when the target program executes these inputs. This is because LLMs cannot fully understand the long texts of input formats or examples due to limited ability on long text processing~\cite{shaham2023zeroscrolls}. For example, Border Gateway Protocol (BGP) is a complex protocol, whose document (BGP RFC 9952) has more than 28,000 words to describe its functionalities. When generating inputs of BGP based on the RFC description, LLMs usually forget to generate the length field of the TLV substructures in the BGP message because the description of the main message structure and the TLV substructures are a little far, making LLMs hard to totally understand BGP format.

\textbf{Recommendations.}  We have the following recommendations:

\recstep Some of the testing inputs to the system are common and have a large number of examples on the web, and they have been included in the \llm{}'s training corpus. \textit{It is possible to directly employ \llm{} to generate test cases for them, combining methodologies focused on diversification.} \label{rec:input1}
These methods encompass internal approaches, such as meticulously crafted prompts that demand using diverse features, as well as external methods, such as coverage-guided genetic algorithms.
They both contribute to address the challenge of \textit{insufficient diversity}.

For instance, when testing common text protocols such as \texttt{HTTP} and \texttt{FTP}, where \llm{} excels in its support for text-based languages, it is feasible to directly instruct \llm{} to generate test cases for these protocols.
To increase diversity, for internal approaches, we can use prompts that encourage \llm{} to generate HTTP files with various methods (e.g., GET, POST, PUT), different headers, different query parameters, URL structures, various payloads, and other aspects. We can also interactively ask \llm{} to cover more types of messages ~\cite{chatafl}.
For external approaches, we can utilize coverage-guided generation used in conventional fuzzing along with more real-world examples to enhance \llm{}.

\recstep In many cases, the \llm{} is not trained with a dedicated training corpus specifically tailored for the test subjects. 
\textit{Rather than employing \llm{} directly for generating the final test cases, we suggest utilizing \llm{} to transform well-known knowledge to formulate the input specifications or build initial test cases.} \label{rec:input2}
The input specification helps address the challenge of \textit{limited validity}, and the initial test cases
help address the challenge of \textit{insufficient diversity}.

For instance, in the case of protocol implementations lacking machine-readable grammar, generating valid test inputs automatically to adhere to the necessary structure and order becomes challenging. 
In such scenarios, leveraging that \llm{} has been trained on established protocols, allows the transfer of grammars from these protocols with the assistance of \llm{} and recorded message sequences.
The grammar can enhance the validity of the generated test cases.
With the grammar, conventional grammar-based fuzzers could be utilized to generate more test cases~\cite{chatafl}.
Another instance is transforming test cases of popular database systems to initial seeds for the tested database system.
The SQL queries of popular database systems like \pg{} have rich diversity and they have already been trained for \llm{}. 
Therefore, leveraging the knowledge of \llm{} to transform them into the format of the target database system is feasible. Providing them to the fuzzer as the initial seed helps enhance the diversity of generated test cases.

\vspace{-0.1cm}
\subsection{Bug Detection} 
\textbf{\indent Description.}
Recently, several pioneer works~\cite{kakarla2023oracle,li2023nuances} utilize LLM to enhance bug detection. Their basic idea is to use functionality descriptions of the target program as the prompt context, and then ask LLMs to generate code that implements the same functionalities with the target program. By comparing the execution results of the two functionally equivalent programs, they can detect logic bugs in the target program.
For example, Differential Prompting~\cite{li2023nuances} queries LLMs about the intention of a piece of provided code and then uses the obtained intention as a new prompt context for LLMs to generate code with the same intention.

\textbf{Challenges.}
The application of LLMs to bug detection can be ineffective if done directly, as LLMs have limited long-text understanding~\cite{shaham2023zeroscrolls}, posing a challenge to \underline{\textit{inaccurate understand}} of the semantics of the target program. 
For example, researchers~\cite{li2023nuances} found that LLMs may misconstrue code designed to identify the longest common substring as being intended for finding the longest common subsequence. This misinterpretation can occur even though these two problems require entirely distinct code solutions. 
As a result, LLMs may generate code whose functionality deviates from the target program, thus leading to an inaccurate test oracle.
According to the experiment results of Differential Prompting~\cite{li2023nuances}, it achieves $66.7\%$ success rate when generating reference implementation for programs from the programming contest website Codeforces. While this is substantially better than its baseline, it still results in a false-positive rate of 33.3\%, which is still not sufficient for practical usage.


\textbf{Recommendations.} We have the following recommendations:

\recstep Defining test oracles is highly dependent on specific targets and scenarios, presenting the most formidable aspect of fuzzing. 
\textit{For complicated targets, we suggest to avoid analyzing results with \llm{} directly. 
\label{rec:oracle1}
Instead, consider employing \llm{} to extract features or patterns associated with a specific bug type, leveraging domain knowledge.} Subsequently, monitoring the system using these patterns aids in addressing the challenge of \textit{inaccurate understanding}. 

For example, many time-series databases like IoTDB implicitly handle exceptions. \
Consequently, the system will not crash or exhibit other abnormal behaviors.
Nevertheless, these database systems generate extensive logs, and errors manifest as exceptions in these logs.
Therefore, it becomes feasible to use \llm{} for analyzing the logs to discern error patterns.
In such scenarios, we recommend employing \llm{} to scrutinize the logs, identify error patterns, and subsequently leverage these patterns for detecting logic errors.

\recstep Some targets or projects contain well-defined documentations, where the expected behaviors are clearly described, like the RFCs for protocols. \emph{For these cases, we suggest to leverage the natural language understanding ability of \llm{} to extract the expected behaviors from the documentations for test oracle definition.}
This helps \llm{} to understand the intention and design of the target programs, thus addressing the challenge of \textit{inaccurate understanding}. \label{rec:oracle2}

For example, the RFCs for protocols usually contain detailed descriptions of the protocol's expected behaviors. Take the RFC 854~\cite{rfc854} for Telnet protocol as an example. It specifies expected behaviors during the negotiation of some disabled command options or unnegotiated commands. These can be used as test oracles and can be further used to uncover CVE-2021-40523~\cite{CVE-2021-40523}.




\section{Potential Solutions}
To demonstrate the practicality of our recommendations, we use the Database Management System (DBMS) as the target for LLM-assisted fuzzing. Addressing challenges in driver synthesis, input generation, and bug detection, we propose three potential solutions: state-aware driver synthesis, cross-DBMS SQL transfer, and log-based Oracle definition.
These solutions are implemented and compared with rudimentary uses of LLM, where it is directly employed. Experiments are conducted under identical settings on a machine with 256 cores (AMD EPYC 7742 Processor @ 2.25 GHz) and 512 GiB of main memory, demonstrating the efficacy of our recommended approaches in enhancing LLM-based fuzzing for intricate systems like DBMSs.

\subsection{LLM-Enhanced Connector Synthesis}
\textbf{Obstacle:} 
Database connectors, also commonly known as database drivers, serve as intermediary components facilitating communication between applications and databases. 
These connectors define standard a set of interfaces, encompassing functions and parameters. 
The driver for fuzzing database connector consists of a sequence of these interfaces.
Directly utilizing \llm{} to generate drivers for database connector will encounter two challenges:
First is \textit{prone to error}: API sequences contain semantic information that is embedded in the state of the database connector, directly generating sequences may import errors. 
Second is \textit{limited scope}: \llm{} lacks the state transition knowledge of the connectors because it lacks the related corpus in training.

\textbf{Solution:} 
Following \ref{rec:driver2}, we propose \llm{}-enhanced state-aware database connector synthesis. 
We first collect JDBC function prototypes and example programs that utilize JDBC.
Then we model the connection relationships between JDBC functions as state-transition rules.
Next, we gather the function prototypes, example programs, and connection rules as input for LLM. 
The prompt we give is like ``
Based on the state-transition rules and state description of functions, please generate a sequence of APIS within length 15. It is required to cover a different combination of state transitions than before.''

\textbf{Result:} 
We implement \llm{}-enhanced connector synthesis into \tool{}$^{conn}$ and compare it against \llm{}$^{conn}$, which directly utilizes \llm{} to generate drivers for MySQL Connector/J~\cite{mysqljgithub}, MariaDB Connector/J~\cite{mariadbjgithub}, and AWS JDBC Driver for MySQL~\cite{awsmysqljgithub}. 
We perform fuzzing on \clickhouse{} for each tool. 
Table~\ref{tab:conn_coverage} shows the driver correctness ratios and branch coverage by LLM$^{conn}$ and \tool{}$^{conn}$ on three selected DBMSs in 12 hours.
These statistics show that \tool{}$^{conn}$ always performs better in both driver correctness ratio and branch coverage than LLM$^{conn}$ on all three DBMSs. 
Specifically,  \tool{}$^{conn}$ archives  94\% more correctness rate for driven synthesis.
And the drivers generated by \tool{}$^{conn}$ cover 56\% more branches on average.
The main reason is that the state-transition rules embed semantic information, and it also helps \llm{} generate API sequences that account for the diverse states within the database connector. 


\begin{table}[htbp]
    \centering
    \caption{Driver Correctness Ratios and Branch Coverage.
    }
    \begin{adjustbox}{width=0.9\linewidth} 
    \begin{tabular}{c|c|c|c|c}
    \toprule
    \multirow{2}{*}{DBMS}
                    & \multicolumn{2}{c|}{Driver Correctness Ratios} &  \multicolumn{2}{c}{Branch Coverage}  \\
    \hhline{~----}
                    & LLM$^{conn}$ & \tool{}$^{conn}$ & LLM$^{conn}$ & \tool{}$^{conn}$    \\
    \hline
         MariaDB Connector/J    & 0.142& 0.331 & 583   & 843  \\
         MySQL Connector/J    &  0.216& 0.367 & 1256   & 1982  \\
        AWS MySQL JDBC &  0.203& 0.394 & 1382  & 2293 \\
         \bottomrule
    \end{tabular}
    \end{adjustbox}
     \label{tab:conn_coverage}
\end{table}

\vspace{-0.2cm}

\subsection{Cross-DBMS SQL Transfer}

\textbf{Obstacle:} 
SQL queries, as the inputs of DBMS, are vital to DBMS fuzzing. Generating SQL queries directly via LLM faces two main challenges: ensuring semantic correctness and promoting query diversity. Semantically correct SQL queries are vital for triggering complex DBMS behaviors, as syntactical errors lead to parsing failures. The intricate SQL grammar, encompassing various clauses, expressions, and rules, poses a challenge for LLM in achieving semantic correct. Furthermore, diversity in SQL queries is crucial for probing deep DBMS logic. However, LLM's constrained variety, influenced by the absence of DBMS feedback, limits the exploration of diverse query structures.


\textbf{Solution:} To overcome these challenges, we introduce the cross-DBMS SQL transfer approach, aligned with the recommendation \ref{rec:input2}, for SQL generation. 
In contrast to directly generating the SQL queries, we use LLM to transfer the test cases from other DBMSs as the initial seeds for fuzzing the target DBMS. These initial seeds are used to mutate new SQL test cases during the fuzzing loop. 
The process contains three key steps. First, it executes existing SQL test cases within its native DBMS to capture the schema information during execution. Second, it utilizes LLMs along with the captured schema information to guide the generation of new test cases based on the LLM responses. Finally, it temporarily comments out unparsable sections for fuzzers to ensure proper parsing and subsequently uncomments them after mutation.

\textbf{Result:}
We implement the solution called \tool{}$^{input}$ and compare it with LLM$^{input}$, which directly uses LLM to generate the SQL queries. 
We run \tool{}$^{input}$ and LLM$^{input}$ on three DBMS: \monetdb{} \cite{monetdb}, \duckdb{} \cite{duckdb}, and \clickhouse{} \cite{clickhouse}. 
\begin{table}[htbp]
    \centering
    \caption{Semantic Correctness Ratios and Branch Coverage.
    }
    \begin{adjustbox}{width=0.85\linewidth} 
    \begin{tabular}{c|c|c|c|c}
    \toprule
    \multirow{2}{*}{DBMS}
                    & \multicolumn{2}{c|}{Semantic Correctness Ratios} &  \multicolumn{2}{c}{Branch Coverage}  \\
    \hhline{~----}
                    & LLM$^{input}$ & \tool{}$^{input}$ & LLM$^{input}$ & \tool{}$^{input}$    \\
    \hline
         \monetdb{}    & 0.1594 & 0.4134 & 26,828   & 41,840  \\
         \duckdb{}     & 0.2551 & 0.3486 & 57,937   & 70,583  \\
         \clickhouse{} & 0.1458 & 0.3093 & 124,887  & 145,383 \\
         \bottomrule
    \end{tabular}
    \end{adjustbox}
    \label{tab:semantic-correctness-compare2}
\end{table}

Table \ref{tab:semantic-correctness-compare2} shows semantic correctness ratios and covered branches of LLM$^{input}$ and \tool{}$^{input}$ on three selected DBMSs in 12 hours. 
From the table, we can see that \tool{}$^{input}$ performs better than LLM$^{input}$ on DBMS fuzzing. 
Specifically, the test cases generated by  \tool{}$^{input}$ contain 159.35\%, 36.65\%, and 112.14\% more semantic-correct SQL statements, and cover 55.96\%, 21.83\%, and 16.41\% more code branches than that of LLM$^{input}$ on \monetdb{},  \duckdb{}, and \clickhouse{}, respectively.
It indicates that LLM can not directly generate high-quality SQL queries as the input for DBMS fuzzing. 
The main reason is that the transfer seeds improve the diversity of mutated test cases, and the fuzzer's mutator promises the semantic correctness of SQL queries.

\vspace{-0.1cm}
\subsection{Monitor-Based DBMS Bug Detection}
\textbf{Obstacle:} The most critical step for DBMS bug detection is to construct the test oracles to identify the logic or performance bugs in DBMS. A test oracle refers to a mechanism in DBMS fuzzing to 
determine the correctness or validity of the DBMS's behaviors. Directly using LLMs to construct the test oracle is challenging as LLMs lack specific knowledge about the intricate workings and behaviors of DBMS. 
They can not access the internal logic, making it difficult to accurately predict or emulate DBMS behavior.

\textbf{Solution:} To address the challenges, we propose the Runtime Monitor-Based DBMS Bug Detection following the \ref{rec:oracle1}, which detects the anomalies of DBMS by analyzing the runtime information of DBMS in real-time.
To ensure the robustness of DBMS, the DBMS usually contains the implicit exception handler mechanism, which captures the internal exceptions to avoid system crashes. These exceptions usually output some key internal states and behaviors of DBMS, such as wrong execution logic.
Unlike directly using LLM to construct the test oracle by checking the execution result of the SQL query, our approach involves collecting runtime information from the DBMS and using LLM to analyze the runtime information for bug detection.
The process contains two main steps. First, it instruments an agent to extract the runtime information of DBMS.
Then, it collects the runtime information and uses LLM to detect the anomaly by predefining some error pattern. 

 \begin{table}[h]
    \centering
    \caption{Number of Reported Bugs and Real Bugs.} \label{tab:bug-detection}
    \setlength{\tabcolsep}{2mm}{
        \scalebox{0.9}[0.9]{%
       \begin{tabular}{l|cc|cc}
            \toprule
                \multicolumn{1}{l|}{DBMS}  &
                \multicolumn{2}{c|}{LLM$^{bug}$} &
                \multicolumn{2}{c}{\tool{}$^{bug}$} 
                \\
                \hline
            \multicolumn{1}{l|}{Name} & \multicolumn{1}{l|}{Reported}  & \multicolumn{1}{l|}{Real}    & \multicolumn{1}{l|}{Reported}  &\multicolumn{1}{l}{Real} \\
\hline
\multicolumn{1}{l|}{\monetdb{}}   & \multicolumn{1}{c|}{61}  & \multicolumn{1}{c|}{0}  &  \multicolumn{1}{c|}{6}             &  \multicolumn{1}{c}{3}      \\

\multicolumn{1}{l|}{\duckdb{}}             & \multicolumn{1}{c|}{54} & \multicolumn{1}{c|}{0} & \multicolumn{1}{c|}{5} & \multicolumn{1}{c}{3}        \\

\multicolumn{1}{l|}{\clickhouse{}}                & \multicolumn{1}{c|}{67}  & \multicolumn{1}{c|}{1} &  \multicolumn{1}{c|}{3}    & \multicolumn{1}{c}{3}              \\



            \bottomrule
            \end{tabular}
        }
    }
\end{table}
\textbf{Result:}
To evaluate the effectiveness of our recommendation, we implement the solution with \tool{}$^{bug}$ and compare it with LLM$^{bug}$, which directly uses LLM to determine whether the execution of the SQL query is right during the fuzz loop. 
Table~\ref{tab:bug-detection} shows the number of reported bugs and real bugs by LLM$^{bug}$ and \tool{}$^{bug}$ in 12 hours on \monetdb{}, \duckdb{} and \clickhouse{}. It shows the \tool{}$^{bug}$ can detect more anomalies  and has fewer false positives than LLM$^{bug}$. Specifically, LLM$^{bug}$ totoally reported 182 bugs but only 1 bug is real. Instead,  \tool{}$^{bug}$ reported 14 bugs and 9 bugs are real bugs and have been confirmed. The main reason is that the collected runtime information contains the error message of DBMS, and it helps LLM to analyze and  detect bugs.

\section{Conclusion}

We identify and systematically analyze five major challenges when using LLM in fuzzing and confirm their prevalence through a review of most recent top-tier conference papers. These challenges affect the effectiveness and accuracy of the LLM-based fuzzing technologies. To support researchers in avoiding them, we provide recommendations that are applicable to effectively assist the main steps in fuzzing, depending on the richness of the relevant corpus and documentation. Our preliminary evaluation further demonstrates that these recommendations effectively address the challenges in LLM-assisted DBMS fuzzing.


\bibliographystyle{ACM-Reference-Format}
\bibliography{FuzzLLM}


\end{document}